\documentstyle{mn}

\include {epsf}

\begin{document}

\def\lesssim{\mathrel{\hbox{\rlap{\hbox{\lower4pt\hbox{$\sim$}}}\hbox{$<$}}}}

\def\gtrsim{\mathrel{\hbox{\rlap{\hbox{\lower4pt\hbox{$\sim$}}}\hbox{$>$}}}}

\def\lsun{{\rm L}_{\odot}}

\def\msun{$M_{\odot}$~}

\title{Evolution  of  DA  white  dwarfs  in  the  context  of
a new theory of convection}

\author[L. G. Althaus and O. G. Benvenuto]
{L. G. Althaus\thanks{Fellow of the Consejo Nacional de
Investigaciones Cient\'{\i}ficas y T\'ecnicas
Argentina.  Email:  althaus@fcaglp.fcaglp.unlp.edu.ar} and
O. G. Benvenuto\thanks{Member  of  the  Carrera  del
Investigador
Cient\'{\i}fico, Comisi\'on  de Investigaciones  Cient\'{\i}ficas
de   la   Provincia   de   Buenos   Aires,   Argentina.    Email:
obenvenuto@fcaglp.fcaglp.unlp.edu.ar}\\
Facultad de Ciencias Astron\'omicas y
Geof\'{\i}sicas, Paseo del Bosque S/N,
(1900) La Plata, Argentina}

\pagerange{206--216}
\volume{296}
\pubyear{1998}

\date{Accepted 1997 November 18. Received 1997 November 18; in
original form 1997 July 8}

\maketitle

\begin{abstract}  In  this  study  we  compute  the structure and
evolution of carbon - oxygen DA (hydrogen - rich envelopes) white
dwarf  models  by  means  of  a detailed and updated evolutionary
code. We consider models with  masses from 0.5 to 1.0  $M_{\sun}$
and we vary the hydrogen layer mass in the interval of  $10^{-13}
\leq  M_{\rm  H}/M  \leq  10^{-4}$.  In  particular, we treat the
energy transport by convection within the formalism of the full -
spectrum turbulence theory,  as given by  the Canuto, Goldman  \&
Mazzitelli (CGM) model. We explore the effect of various hydrogen
layer  masses  on  both  the  surface  gravity  and  the hydrogen
burning.   Convective   mixing   at   low  luminosities  is  also
considered.

One of  our main  interests in  this work  has been  to study the
evolution of ZZ  Ceti models, with  the aim of  comparing the CGM
and mixing - length theory (MLT) predictions. In this connection,
we find that  the temperature profile  given by the  CGM model is
markedly different from that of  the ML1 and ML2 versions  of the
MLT.  In  addition,  the  evolving  outer convection zone behaves
differently in both theories.

We have also computed approximate effective temperatures for  the
theoretical blue edge of DA instability strip by using thermal  -
timescale arguments for our evolving DA models. In this  context,
we found that the CGM theory leads to blue edges that are  cooler
than the  observed ones.  However, because  the determination  of
atmospheric  parameters  of  ZZ  Ceti  stars  is dependent on the
assumed convection description in model atmosphere  calculations,
observed  blue   edges  based   on  model   atmospheres  computed
considering the  CGM theory  are required  in order  to perform a
self -  consistent comparison  of our  results with observations.
Finally, detailed non - adiabatic pulsational computations of  ZZ
Ceti models considering the CGM convection would be necessary  to
place the results found in this paper on a firmer basis.

\end{abstract}


\section{introduction}

White  dwarf  stars  with  hydrogen  lines  (spectral  type   DA)
constitute the most numerous group of the observed white  dwarfs.
Since  the  first  DA   white  dwarf  was  reported   to  exhibit
multiperiodic  luminosity  variations  (see  McGraw  1979),   the
interest in these objects  has greatly increased. Rapid  progress
in this  field has  been possible  thanks to  the development  of
powerful theoretical tools as well  as to a increasing degree  of
sophistication  in  observational  techniques.  Over  the  years,
various studies have presented strong evidence that pulsating  DA
white dwarfs  (DAV) or  ZZ Ceti  stars represent  an evolutionary
stage in the cooling history of  the majority of, if not all,  DA
white dwarfs (Weidemann \& Koester 1984 and references therein).

Studies  carried  out  notably  by  Dolez  \& Vauclair (1981) and
Winget  et  al.  (1982)  on  the  basis  of  detailed
non - radial,
non - adiabatic  pulsation   calculations,  demonstrated   that
the
$\kappa$-mechanism operating in  the hydrogen partial  ionization
zone is responsible for  the g(gravity)-mode instabilities in  ZZ
Ceti stars. From  then on, pulsating  white dwarfs have  captured
the  interest  of  numerous  investigators, who have employed the
powerful   technique   of   white   dwarf   seismology  to derive
fundamental parameters of these stars, such as the stellar  mass,
chemical composition and stratification of the outer layers (see,
for example, Winget et al.  1994 and Bradley 1996 in  the context
of   DB   and   DA,   respectively).   In   particular, pulsation
calculations have shown that the theoretical determination of the
location of the blue (hot)  edge of the instability strip  (which
marks  the  onset  of  pulsations)  in  the Hertzsprung - Russell
diagram is  mostly sensitive  to the  treatment of  convection in
envelope models. Accordingly, a  comparison with the location  of
the hottest observed ZZ Ceti  stars allows us to obtain  valuable
information  about  the  efficiency  of  convection in such stars
(Winget et al. 1982; Tassoul, Fontaine \& Winget 1990; Bradley \&
Winget 1994; Bradley 1996 and references cited therein).

Unfortunately, the determination of atmospheric parameters of  ZZ
Ceti  stars  is  strongly  dependent  on  the  assumed convection
description in model  atmosphere calculations, against  which the
observed spectrum is  compared. The theoretical  analysis carried
out  by  Bergeron,  Wesemael  \&  Fontaine (1992a) is
particularly
noteworthy in  this regard.  In fact,  these authors demonstrated
that the predicted emergent fluxes, color indices and equivalent
widths of DA white dwarfs, notably of the ZZ Ceti stars, are very
sensitive to the details of  the parametrization of the mixing  -
length theory (MLT) of convection. In particular, Bergeron et al.
(1992a)  concluded  that  the  uncertainties  in the
observational
definition of  the blue  edge of  the ZZ  Ceti instability  strip
brought  about  by  different  convective  efficiencies  may   be
appreciable, irrespective  of the  observational technique that
is used.

Needless  to  say,  the  employment  of   MLT  to  deal  with
convection  represents  a  serious  shortcoming  in   theoretical
studies of white dwarfs. In particular, MLT approximates  the
whole  spectrum  of  eddies  necessary  to  describe  the stellar
interiors by a single large eddy. This representation  introduces
in the description of the model certain free parameters (B\"{o}hm
- Vitense 1958) which, in most stellar studies, are reduced to  a
single  one:  the  distance  $l$  travelled  by  the  eddy before
thermalizing with the surrounding  medium. $l$ is written  as $l=
\alpha H_{\rm p}$ where $H_{\rm p}$ is the pressure scaleheight
and $\alpha$ is the  free parameter. Obviously, the  existence of
such an adjustable parameter prevents stellar evolution
calculations
from being completely predictive. In white dwarf studies, several
parametrizations   of   the   MLT   are   extensively   employed.
Specifically, the ML1, ML2  and ML3 versions, which  differ
in  the  choice  of  the  coefficients  appearing  in   the
expressions of the convective flux, average velocity and
convective efficiency (see Tassoul  et al. 1990 for  details) are
associated, respectively,  with increasing  convective efficiency
(for the ML1 and ML2 versions $\alpha = 1$, while  ML3 is  the
same as ML2 but with $\alpha= 2$).

Over  the  last  two  decades,  an  unparalleled  effort has been
devoted to the observational determination of the location of the
blue edge  of the  instability strip  of the  ZZ Ceti  stars (see
Wesemael  et  al.  1991   for  a  review).  Based   on  different
observational techniques,  the majority  of the  studies cited by
Wesemael et  al. point  towards a  temperature of  the blue edge,
defined by the ZZ Ceti star G117-B15A, of approximately 13000  K.
New observational  data and  improved model  atmospheres tend  to
indicate a significantly lower temperature for the blue edge.  In
fact,  from  ultraviolet  observations  performed with the Hubble
Space Telescope,  Koester, Allard  \& Vauclair  (1994) have shown
that model  atmospheres calculated  with the  ML1 parametrization
do  not  fit  satisfactorily  to  the  ultraviolet  spectrum of
G117-B15A. Instead, a more efficient convection, as given by the
ML2 version, is required to explain the observed spectrum. On the
basis of  this result,  Koester et  al. derived  for G117-B15A an
effective temperature of 12250 $\pm$  125 K. In view of  the fact
that the  ultraviolet spectra  of other  candidates for  the blue
edge of  the ZZ  Ceti instability  strip are  similar to  that of
G117-B15A,  Koester   et  al.   concluded  that   the  blue  edge
temperature  is  likely  to  be  around  12250  K, a value which
is considerably lower than found in previous studies.

In a still more recent analysis, Bergeron et al. (1995)  inferred
an  even   lower  effective   temperature  for   G117-B15A.  More
precisely, on the basis  of new optical and  ultraviolet analyses
of ZZ Ceti stars, Bergeron  et al. succeeded in constraining  the
convective efficiency  used in  model atmosphere  calculations of
these  stars.  Indeed,  they  demonstrated that a parametrization
less  efficient  than  ML2,  particularly  the   ML2/$\alpha$=0.6
version (intermediate in efficiency  between ML1 and ML2)  yields
ultraviolet temperatures that are completely consistent  with the
optical  determinations.  In  this  respect,  they  derived   for
G117-B15A  an  effective  temperature  of  11620 K. In the sample
analysed by  Bergeron et  al., the  blue edge  of the instability
strip is defined by the  star G226-29, for which they  obtained a
temperature of 12460 K and  a surface gravity of $\log  g= 8.29$.
Finally, Fontaine  et al.  (1996) presented  a reanalysis  of the
observed  pulsation  modes  of  G117-B15A  that  reinforces   the
conclusions  of  Bergeron  et  al.  (1995)  on  the   atmospheric
parameters  of  this  variable  white  dwarf.  Fontaine  et   al.
concluded that a model  atmosphere of G117-B15A characterized  by
an   effective   temperature   of   $\approx$   11500   K   and a
ML2/$\alpha$=0.6  convection  is  consistent  with  optical   and
ultraviolet  spectroscopic  observations  as  well  as  with  the
observed pulsation amplitudes in different wavelength bands.

Another  conclusion  drawn  by  Bergeron  et al. (1995) (see also
Wesemael et al.  1991) worthy of  comment is related  to the fact
that convective efficiency in the upper atmosphere appears to  be
quite  different  from  that  in  the  deeper  envelope. In fact,
according  to  non - adiabatic  pulsation  calculations  (Bradley
\&
Winget  1994  and  references  cited  therein),  a more efficient
convection than  that provided  by ML2/$\alpha$=0.6  is needed in
the deeper envelope  to explain the  observed DA blue  edge. This
result,  which  reflects  the   inability  of  MLT  to   describe
convection   throughout   the   entire   outer   convection  zone
adequately,  has  indeed  been  borne  out  by  recent   detailed
hydrodynamical simulations  of convection  in ZZ  Ceti stars (see
later in this section). Other objections have been raised against
MLT.  For  instance,  the  value  of  $\alpha$,  which is usually
derived from solar radius  adjustments, is larger than  unity, in
contrast to the basic postulates  of MLT. In addition to  this,
the procedure of applying  the solar $\alpha$ parameter  to other
stars is not  entirely satisfactory, at  least in the  context of
red giant  stars for  which a  wide range  of $\alpha$  values is
needed (Stothers \& Chin 1995,  1997). Finally, MLT does not  fit
laboratory convection data (Canuto 1996).

Fortunately, a  considerable effort  has been  devoted in  recent
years  to  improving  some  of  the  basic  postulates of MLT. In
particular, Canuto \& Mazzitelli  (1991, 1992, hereafter CM)  and
more  recently,  Canuto,  Goldman  \& Mazzitelli (1996, hereafter
CGM) developed two convection models based on the full - spectrum
turbulence (FST) theory,  which represent a  substantial progress
in modelling stellar convection (we refer the reader to Canuto \&
Christensen - Dalsgaard 1997 for a recent review on local and non
-  local  convection  models).  Both  of  the  treatments resolve
essentially two main shortcomings of the MLT

\indent (i) The  MLT one -  eddy approach is  replaced by a  full
spectrum  of  turbulent  eddies  derived  on  the  basis of amply
tested, modern theories  of turbulence. As  discussed by CM,  the
one - eddy approximation is only justified for viscous fluids but
becomes  completely  inadequate  for  almost - inviscid stellar
interiors.

\indent (ii) There are
no adjustable parameters.

These features render  the CM and  CGM models substantially  more
solid  and  complete  than  MLT.  The  CM model in particular has
successfully   passed   a   wide   variety   of   laboratory  and
astrophysical tests,  performing in  all cases  much better  than
MLT. With regard to the astrophysical case, the CM model has been
applied to the study of different kind of stars and  evolutionary
phases.  Applications  include  the  solar  model  (see  CM), the
effective temperature of which is fitted within $\approx 0.5$ per
cent accuracy; red giants  (D'Antona, Mazzitelli \& Gratton  1992
and  Stothers  \&  Chin  1995,  1997),  for which MLT may require
various values of  $\alpha$ for different  stars, helioseismology
(Patern\`o  et  al.  1993  and Monteiro, Christensen-Dalsgaard \&
Thompson  1996  amongst  others),  low  - mass stars (D'Antona \&
Mazzitelli 1994)  and stellar  atmospheres (Kupka  1996). In  the
white dwarf context,  the CM model  has been shown  to be a  very
valuable tool as well. In fact, Althaus \& Benvenuto (1996)  (see
also Mazzitelli \& D'Antona 1991) studied the evolution of carbon
- oxygen white dwarfs with helium - rich envelopes (spectral type
DB) and found that the  CM model predicts theoretical blue  edges
for  the  DB  instability  strip  in  good  agreement  with   the
observations  of  Thejll,  Vennes  \&  Shipman (1991). Mazzitelli
(1995) has also applied the  CM model to the study  of convection
in DA white dwarfs.  More recently, Althaus \&  Benvenuto (1997a)
used the CM model to  study the evolution of helium  white dwarfs
of low and intermediate masses.

CGM  have  recently  improved  the  CM  model  by  developing   a
self-consistent model  for stellar  turbulent convection,  which,
like  the  CM  model,  includes  the  full spectrum of eddies but
computes  the  rate  of  energy  input  self - consistently. More
precisely, CGM  resolve analytically  a full  turbulence model in
the local limit, taking  into account  the fact  that the  energy
input from the source  (buoyancy) into the turbulence  depends on
both  the  source   and  the  turbulence   itself.  At  low   and
intermediate  convective  efficiencies,  the  CGM  gives  rise to
higher convective fluxes  than those given  by the CM  model. CGM
find that the main sequence  evolution of a solar model  does not
differ appreciably from the CM predictions but demands a  smaller
overshooting  to  fit  the  solar  radius,  in agreement with new
observational  data.  D'Antona,  Caloi  \& Mazzitelli (1997) have
also  applied  the  CGM  model  to  study the problem of globular
cluster ages. In the white dwarf domain, the CGM model also fares
better  than  the  CM  model  (Althaus  \&  Benvenuto  1997b  and
Benvenuto  \&  Althaus  1997).  Finally,  Benvenuto  \&  Althaus
(1997b) have applied the CGM model to study the evolution of  low
-  and  intermediate  -  mass  helium  white dwarfs with hydrogen
envelopes.

In order to assess  to what extent the  MLT is able to  provide a
reliable  description  of  convection  in  the outer layers of DA
white dwarfs,  Ludwig, Jordan  \& Steffen  (1994) have  performed
detailed  two   -  dimensional   hydrodynamical  simulations   of
convection at the  surface of a  ZZ Ceti star.  They demonstrated
that the  temperature profile  in the  outer layers  of a $T_{\rm
eff}$ = 12600  K, $\log g$  = 8.0 DA  model cannot be  reproduced
with model atmospheres calculated with  MLT and with a  single
value of $\alpha$. Instead, the value of $\alpha$ is to be varied
from 1.5 in the  upper atmosphere to 4  in much deeper layers  in
order for the hydrodynamical temperature profile to be reasonably
well represented.  Another result  obtained by  Ludwig et  al. is
that their  hydrodynamical simulations  lead to  convective zones
that are  too shallow  for the  $\kappa$, $\gamma$  mechanism to
give rise unstable  g modes,  implying a blue  edge for the  ZZ
Ceti instability  strip significantly  cooler than  12600 K. This
important finding has recently been borne out by Gautschy, Ludwig
\&  Freytag  (1996)  on  the  basis  of  non - radial,
non - adiabatic
pulsation  calculations.  Gautschy  et  al.  included  in   their
analysis Ludwig et al.'s hydrodynamical simulations to derive
the  structure  of  the  convective  layers  of their white dwarf
models. This feature  represents  substantial  progress compared
with earlier investigations,  which employ  MLT to deal  with
convection. Gautschy et al. are unable to find a theoretical blue
edge  for  the  ZZ  Ceti  instability  strip  consistent with the
observations  of  Bergeron  et  al.  (1995)  and  conclude  that,
according  to  their  calculations,  it  lies between 11400 K and
11800 K (for a model with $\log g= 8$).

The present study  is devoted to  presenting new calculations  of
the evolution of carbon - oxygen DA white dwarfs considering  the
model for  stellar turbulent  convection of  CGM as  well as  the
various parametrizations  of MLT  commonly used  in the  study of
these stars.  Attention is  focused mainly  on those evolutionary
stages corresponding to the ZZ Ceti effective temperature  range.
The  calculations  are  carried  out  with a detailed white dwarf
evolutionary code in which we include updated equations of state,
opacities and neutrino emission rates. Furthermore, we take  into
account crystallization, convective mixing and hydrogen  burning.
To explore the sensitivity of the results to various input  model
parameters,  we  vary  both  the  model  mass  from  0.50 - 1.0
$M_{\sun}$, which covers most of the observed mass distribution
of
DA white dwarfs (Bergeron, Saffer \& Liebert 1992b, Marsh et  al.
1997), and the  hydrogen layer mass  $M_{\rm H}$ in  the interval
$10^{-13} \leq M_{\rm H}/M \leq  10^{-4}$ where $M$ is the  model
mass. Finally the hydrogen - helium transition zone is assumed to
be almost discontinuos. Details of our evolutionary code and  its
main physical constituents are presented in Section 2. Section  3
is devoted to analysing our results and, finally, in Section 4 we
summarize our findings.

\section{computational details and input physics}

The white dwarf evolutionary code  we employed in this study  was
developed by  us independently  of other  researchers and  it has
been used to  study different problems  connected to white  dwarf
evolution as well as the cooling of the so - called strange dwarf
stars (Benvenuto \& Althaus 1996). Details of the code and its
constitutive physics can be found in Althaus \& Benvenuto (1997a)
and Benvenuto \& Althaus  (1997, 1998). In what follows  we
restrict ourselves to a few brief comments.

The equation of  state for the  low - density  regime is that  of
Saumon,  Chabrier  \&  Van  Horn  (1995)  for hydrogen and helium
plasmas. The treatment for the high - density, completely ionized
regime appropriate  for the  white dwarf  interior includes ionic
and   photon   contributions,   coulomb  interactions,  partially
degenerate electrons  and electron  exchange and  Thomas -  Fermi
contributions at finite temperature. Radiative opacitites for the
high -  temperature regime  ($T \geq  8000$ K)  are those of OPAL
(Iglesias \& Rogers 1993) with metallicity $Z=0.001$, whilst  for
lower  temperatures  we  use  the  Alexander  \&  Ferguson (1994)
molecular  opacities.  Molecular   effects  become  relevant   at
temperatures as high as 5000 K and below 2500 K they are  dominat
(Alexander \& Ferguson 1994). We shall see that convective mixing
at low effective temperatures may in some cases change the  outer
layer  composition  from  almost  pure  hydrogen  to  almost pure
helium.  This  being  the  case,  we  had  to  rely, in the low -
temperature regime,  on the  older tabulation  of Cox  \& Stewart
(1970) for helium composition. Conductive opacities for the low -
density regime are taken from  Hubbard \& Lampe (1969) as  fitted
by  Fontaine  \&  Van  Horn  (1976). Conductive opacities for the
liquid  and  crystalline  phases,  and  the various mechanisms of
neutrino emission  relevant to  white dwarf  interiors are  taken
from  the  works  of  Itoh  and  collaborators  (see  Althaus  \&
Benvenuto  1997a  and  references  cited  therein).  It  is worth
mentioning  that  neutrino  cooling  is  dominant  during the hot
phases  of  white  dwarf  evolution  but  they  become completely
negligible at the ZZ Ceti effective temperature range.

The  procedure  we  followed  to  generate  the initial models of
different masses  is described  in Benvenuto  \& Althaus  (1995).
Such  initial  models  were  constructed  starting  from  a $0.55
M_{\sun}$ carbon - oxygen white dwarf model kindly provided to us
by Professor Francesca  D'Antona. We assumed  the same core
chemical
composition (consisting of  a mixture of  carbon and oxygen,  see
Fig.  1  of  Benvenuto  \&   Althaus  1995)  for  all  of   them,
notwithstanding the changes in the internal chemical  composition
that  are  expected  to  occur  for  models with different masses
because of  differences in  the pre  - white  dwarf evolution  of
progenitors objects. The  carbon - oxygen  core of our  models is
surrounded by an almost pure helium envelope, the mass of which
($M_{\rm
He}$) was  varied in  the range  $10^{-6} \leq  M_{\rm He}/M \leq
10^{-2}$. In  DA white  dwarfs there  is an  almost pure hydrogen
envelope on the top of the helium layers. Unfortunately, the mass
of this  hydrogen envelope  is only  weakly constrained  by pre -
white  dwarf  evolutionary  calculations  (D'Antona \& Mazzitelli
1991)  and  by  non - adiabatic  pulsation  studies (Fontaine et
al.
1994).  In  recent  years,  however,  strong  evidence  has  been
accumulated in favour of the idea that some ZZ Ceti stars  appear
to have  thick hydrogen  layers (Fontaine  et al.  1994). In  the
present  study,  we  decided  to  treat  the mass of the hydrogen
envelope as basically a free parameter within the range $10^{-13}
\leq M_{\rm H}/M \leq 10^{-4}$. We refer the reader to  Benvenuto
\& Althaus (1998) for details  about the procedure we follow  to
add a  hydrogen envelope  at the  top of  our models.  We want to
mention that we assumed the hydrogen/helium transition zone to be
almost discontinuous.

We also included in our evolutionary calculations the release  of
latent  heat  during  crystallization  (see  Benvenuto \& Althaus
1997) and the effect of  the convective mixing. We shall  see in
particular that the onset of crystallization in our more  massive
models occurs  at effective  temperatures close  to the  observed
blue edge. Finally,  we also considered  the effects of  hydrogen
burning by including in  our numerical code the  complete network
of   thermonuclear   reaction   rates   for   hydrogen    burning
corresponding  to  the  proton  -  proton  chain and the CNO bi -
cycle. Nuclear reaction rates  are taken from Caughlan  \& Fowler
(1988) and $\beta$ - decay rates from Wagoner (1969), taking into
account the corrections for their Q - values resulting from the
effect of
neutrino losses. Electron screening  is from Wallace, Woosley  \&
Weaver  (1982).  We  consider  the  following  chemical  species:
$^{1}$H,   $^{2}$H,   $^{3}$He,   $^{4}$He,  $^{7}$Li,  $^{7}$Be,
$^{8}$B,  $^{12}$C,   $^{13}$C,  $^{13}$N,   $^{14}$N,  $^{15}$N,
$^{15}$O, $^{16}$O, $^{17}$O and $^{17}$F.

The most relevant feature of the present study is the  employment
of the FST  theory to deal  with energy transport  by convection.
This  theory  represents  a  substantial  improvement  over  MLT,
particularly in the treatment of low - viscosity fluids like  the
ones present in stellar interiors,  for which the MLT one  - eddy
assumption  is  completely  inadequate.  In  particular, a self -
consistent model without adjustable  parameters based on the  FST
approach  has  been  recently  formulated  by  CGM. These authors
fitted their theoretical results for the convective flux  $F_{\rm
c}$ with the expression

\begin{equation}
F_{\rm c}= K T H_{\rm p}^{-1} (\nabla - \nabla_{\rm ad})
\Phi,
\end{equation}

\noindent  where  $K=  4acT^3/3\kappa  \rho$  is  the   radiative
conductivity,  $H_{\rm  p}$  is  the  pressure  scaleheight,  and
$\nabla$ and  $\nabla_{\rm ad}$  are, respectively,  the true and
adiabatic temperature gradients. $\Phi$ is given by

\begin{equation} \Phi= F_{\rm 1}(S) F_{\rm 2}(S),
\end{equation}

\noindent where

\begin{equation} F_1(S)= ({{Ko}\over{1.5}})^{3}
aS^{k}[(1+bS)^{l}-1]^{q},  \end{equation}

\noindent and

\begin{equation}
F_{\rm 2}(S)=
1+{{cS^{0.72}}\over{1+dS^{0.92}}}+{{eS^{1.2}}\over{1+fS^{1.5}}}.
\end{equation}

\noindent Here, $Ko$  is the Kolmogorov constant  (assumed
to be 1.8), $S= 162\ A^{2}(\nabla - \nabla_{\rm ad})$,
and the coefficients are given by $a=10.8654$, $b=4.89073  \times
10^{-3}$,  $k=0.149888$,  $l=0.189238$,  $q=1.85011$,  $c=1.08071
\times 10^{-2}$, $d= 3.01208 \times 10^{-3}$, $e= 3.34441  \times
10^{-4}$, and $f= 1.25 \times 10^{-4}$. $A$ is given by

\begin{equation}        A=        {{c_{\rm       p}\rho^{2}\kappa
z^2}\over{12acT^{3}}}  {({{g  \delta}\over{2H_{\rm  p}}})^{1/2}}.
\end{equation}

A important  characteristic of  the CGM  model is  the absence of
free parameters.  In particular,  the mixing  length is  taken as
$l=z$ where $z$ is the  geometrical distance from the top  of the
convection zone. Stothers \& Chin (1997) have recently shown that
$l=z$ performs very well in different kind of stars. In order  to
avoid  numerical  difficulties, we  allowed  for  a  very   small
overshooting  at  the  moment  of  evaluating  $l$,  that  is, we
considered  $l=  z  +  \beta  H_{\rm  p}^{top}$,  where   $H_{\rm
p}^{top}$ is the  pressure scaleheight  at the top  of the outer
convection zone. We elected $\beta < 0.1$, which does not  affect
the temperature profile  of the convection  zone of our  evolving
models.

For the sake of comparison, we also included in our  calculations
the ML1, ML2 and ML3 parametrizations of MLT. These versions,
which are associated with different convective efficiencies,  has
been amply employed in white dwarf investigations (see Tassoul at
al. 1990 for details).

\section{evolutionary results}

Here,  we  present  the  main  results  of  our  calculations. We
computed  the  evolution  of  DA  white  dwarf models with masses
ranging from $M= 0.5M_{\sun}$ to $M= 1.0M_{\sun}$ at intervals of
$0.1M_{\sun}$ and  with a  metallicity of  $Z= 0.001$.  We varied
$M_{\rm H}$ within  the range $10^{-13}  \leq M_{\rm H}/M  \leq 2
\times  10^{-4}$  and  $M_{\rm  He}$  in  the range $10^{-6} \leq
M_{\rm He}/M \leq 10^{-2}$. We  employ the FST approach given  by
CGM, and  also the  ML1, ML2  and ML3  versions of  MLT.  The
models  were  evolved  from  the  hot  white  dwarf stage down to
$\log{(L/L_{\sun})}  =  -5$.  Unless  stated  otherwise, we shall
refer in what follows to those models parametrized by $M_{\rm H}=
10^{-6}M$ and $M_{\rm He}= 10^{-2}M$.

We begin by  examining Fig. 1  in which the  surface
gravity of our DA models with $\log q(\rm H)= -10, -6, -4$ and no
hydrogen  envelope  is  depicted   in  terms  of  the   effective
temperature ($q$ is the external  mass fraction given by $q=  1 -
M_{r} /  M$)\footnote{For the  sake of  clarity, in  this and the
following figures, we do not depict the results corresponding  to
1.0 $M_{\sun}$ models}.  The first observation  we can make  from
this figure is that  thick hydrogen envelopes appreciably  modify
the surface gravity  of no -  hydrogen models, especially  in the
case of low - mass configurations. The effect of adding an  outer
hydrogen envelope  of $M_{H}=  10^{-10}$ M  is barely noticeable.
Note also that finite - temperature effects are relevant even  in
massive models (see also Koester \& Sch\"{o}nberner 1986). At low
effective temperatures,  convective mixing  between hydrogen  and
helium  layers  (see  later  in  this section) change the surface
gravity  of  models  with   thin  hydrogen  envelopes  (see   the
discussion in Benvenuto \& Althaus 1998 in the context of low  -
mass helium white dwarfs for more details about this topic).

It is worthwhile to mention that our more massive models begin to
develop a crystalline core at effective temperatures close to the
temperature  range  of  the  ZZ  Ceti  instability  strip.   More
specifically, the onset of crystallization at the centre of  1.0,
0.9, 0.8 and 0.7  $M_{\sun}$ DA models  with $\log q= -6$  takes
place  at  $T_{\rm  eff}=$  16200,  13100,  10870  and  8900   K,
respectively. For details concerning the process of growth of the
crystal phase of our models, see Benvenuto \& Althaus (1995).

\begin{figure}
\epsfxsize=210pt
\begin{displaymath}
\epsfbox{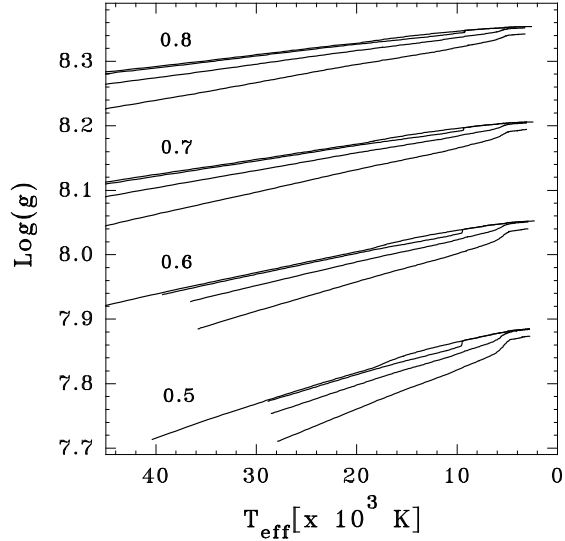}
\end{displaymath}
\caption{
Surface gravity  versus the  effective
temperature for DA white dwarf  models with (from top to  bottom)
$M/M_{\sun}=$ 0.8, 0.7, 0.6 and  0.5. For each stellar mass,  the
four curves correspond  to models with  (from top to  bottom): no
hydrogen envelope, and $\log q(\rm H)$= -10, -6 and -4. Note that
very  thick  hydrogen  envelopes  appreciably  modify the surface
gravity   of   the   no   -   hydrogen  models.
}
\end{figure}

\begin{figure}
\epsfxsize=210pt
\begin{displaymath}
\epsfbox{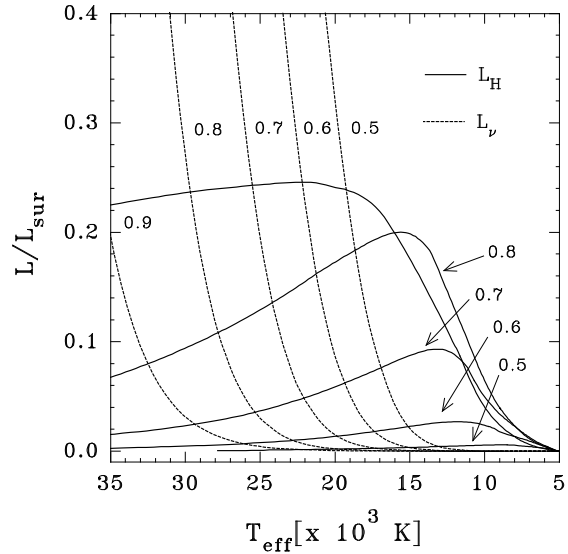}
\end{displaymath}
\caption{
Hydrogen   -   burning (solid lines)  and   neutrino
luminosity (dotted lines), (in  terms  of  the  surface   photon
luminosity) as a function  of the effective temperature  for 0.5,
0.6, 0.7, 0.8 and 0.9 $M_{\sun}$ DA white dwarf models with $\log
q(\rm H)= -4$.
}
\end{figure}

The role  played by  hydrogen burning  in our  evolving models is
worthy  of  comment.  Iben  \&  Tutukov  (1984) were the first in
drawing  the  attention  to  the  fact  that  hydrogen burning in
cooling  white  dwarfs  could  be,  at  low luminosities, a major
energy source. We find in  particular that for models with  $\log
q(\rm  H)  \gtrsim  -4$,  hydrogen  burning  may contribute non -
negligibly to the surface luminosity. For more details, we depict
in Fig.  2 the  relative contribution  of hydrogen  burning as  a
function  of  effective  temperature  for  various stellar masses
having a hydrogen envelope of $\log q(\rm H)= -4$. Note that  for
intermediate  stellar  masses the  hydrogen  -  burning
contribution
reaches a  maximum near  the ZZ  Ceti instability  strip. Another
feature worthy of mention shown by Fig. 2 is that, even for  more
massive  models,  hydrogen  burning  never  becomes  the dominant
source of stellar energy. We also include in Fig. 2 the  neutrino
luminosity of  the models.  For the  effective temperature  range
shown in the figure, neutrino luminosity strongly decreases  with
cooling  and  it  is  negligible  by  the  time  models enter the
instablitiy strip.

The importance  of nuclear burning  is strongly sensitive  to
the excat value of the hydrogen layer mass, as was recognised  by
Koester  \&   Sch\"{o}nberner  (1986)   (see  also   D'Antona  \&
Mazzitelli  1979).  To  show  this,  we  have computed additional
sequences by considering the cases with $\log q(\rm H)=$  -3.824;
-3.699.  The  results  for  0.6  and  0.8  $M_{\sun}$  models are
illustrated in  Fig. 3.  It is  clear that  a small
difference  in  the  hydrogen  layer  mass is responsible for the
different role of  hydrogen burning. In  particular, for the  0.6
$M_{\sun}$ model with $M_{\rm H}= 9 \times 10^{-5} M_{\sun}$,  we
find that the  relative contribution of  hydrogen burning at  low
luminosities remains always below 9 per cent, while for the same
model
but  with  $M_{\rm  H}=  1.2  \times  10^{-4} M_{\sun}$,
the hydrogen - burning contribution rises up to $\approx 18$ per
cent

The behaviour of the  evolving outer convection zone  is depicted
in Figs. 4 - 7 for  DA white  dwarf
models with  $M=$ 0.5,  0.6, 0.8  and 1.0  $M_{\sun}$ and for the
different theories  of convection.  In each  figure, we  plot the
location of the top and the base of the outer convection zone  in
terms  of  the  mass  fraction  $q$  as  a  function of effective
temperature. The first observation we can make from these figures
is  that  at  both  hot  and  cool  extremes the convection zone
profile is independent of  the treatment of convection.  In fact,
at high  effective temperatures  only a  negligible fraction  of
energy   is   carried   by   convection   and,  consequently, the
temperature stratification  remains essentially  radiative, while
at low temperatures convection is very efficient and most of  the
convection   zone   assumes   an   adiabatic   stratification. At
intermediate effective  temperatures, however,  where the  ZZ
Ceti
stars are  observed, convection  treatment is  decisive in fixing
the structure  of the  outer zone  of the  models. Note  that, in
contrast to the situation  encountered for DB objects  (Benvenuto
\& Althaus 1997), the  CGM predictions are roughly  intermediate
to those of the ML1 and ML2 convection. Another feature shown  by
these figures and worthy of
\begin{figure}
\epsfxsize=210pt
\begin{displaymath}
\epsfbox{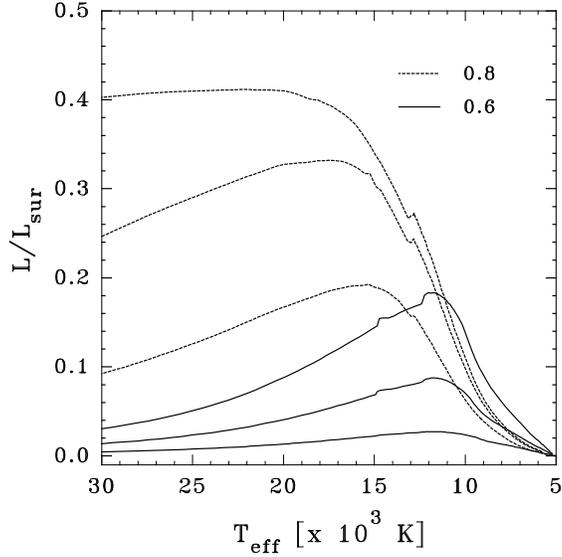}
\end{displaymath}
\caption{
Hydrogen -  burning luminosity (in  terms
of the surface photon luminosity) as a function of the  effective
temperature for 0.6 and 0.8 $M_{\sun}$ DA white dwarf models. For
each stellar mass and from top to bottom, models with $\log q(\rm
H) =  $ -3.699,  -3.824 and  -4 are  depicted.
}
\end{figure}
comment is that the thickness of  the
convection zone in  the CGM model  begins to increase  at a given
effective temperature much  more steeply than  in any of  the MLT
versions. A similar trend is also found in DB white dwarf models,
though it  is worth  remarking that  the global  behaviour of the
evolving convection zone is markedly different in both  types
of  stars.  In  agreement  with  previous  results  (D'Antona  \&
Mazzitelli 1979),  the final  extent reached  by the  base of the
convection zone is smaller in the more massive models. We should
mention that the small jump at the top of the convection zone  at
$T_{\rm eff} \approx 9000 K$ is brought about, not  surprisingly,
by  the  change  from  OPAL  to Alexander \& Ferguson's
(1994) molecular
opacities. This jump affects the $z$ values, thus giving
rise to some  irregularities in  the CGM  convection profile,
which does not alter the conclusions of this work.
\begin{figure}
\epsfxsize=210pt
\begin{displaymath}
\epsfbox{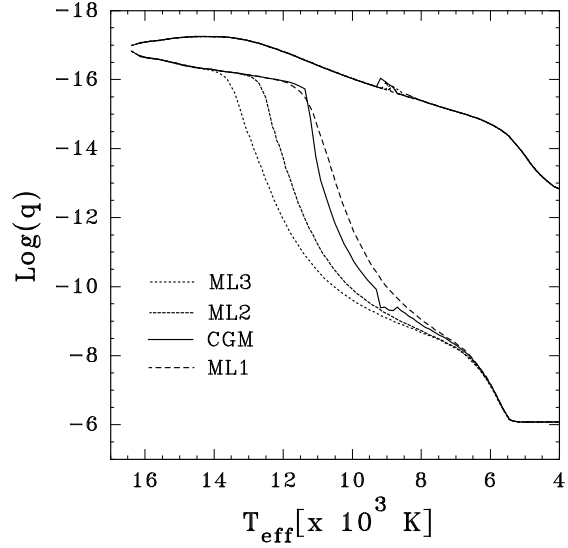}
\end{displaymath}
\caption{
The location of  the top and the  base of
the outer convection zone expressed in terms of the mass fraction
$q$  versus  the  effective  temperature  for a 0.6 $M_{\sun}$ DA
white dwarf model according to different theories of  convection.
For the  small jump  at $T_{\rm  eff} \approx  9000$ K see text.
}
\end{figure}

\begin{figure}
\epsfxsize=210pt
\begin{displaymath}
\epsfbox{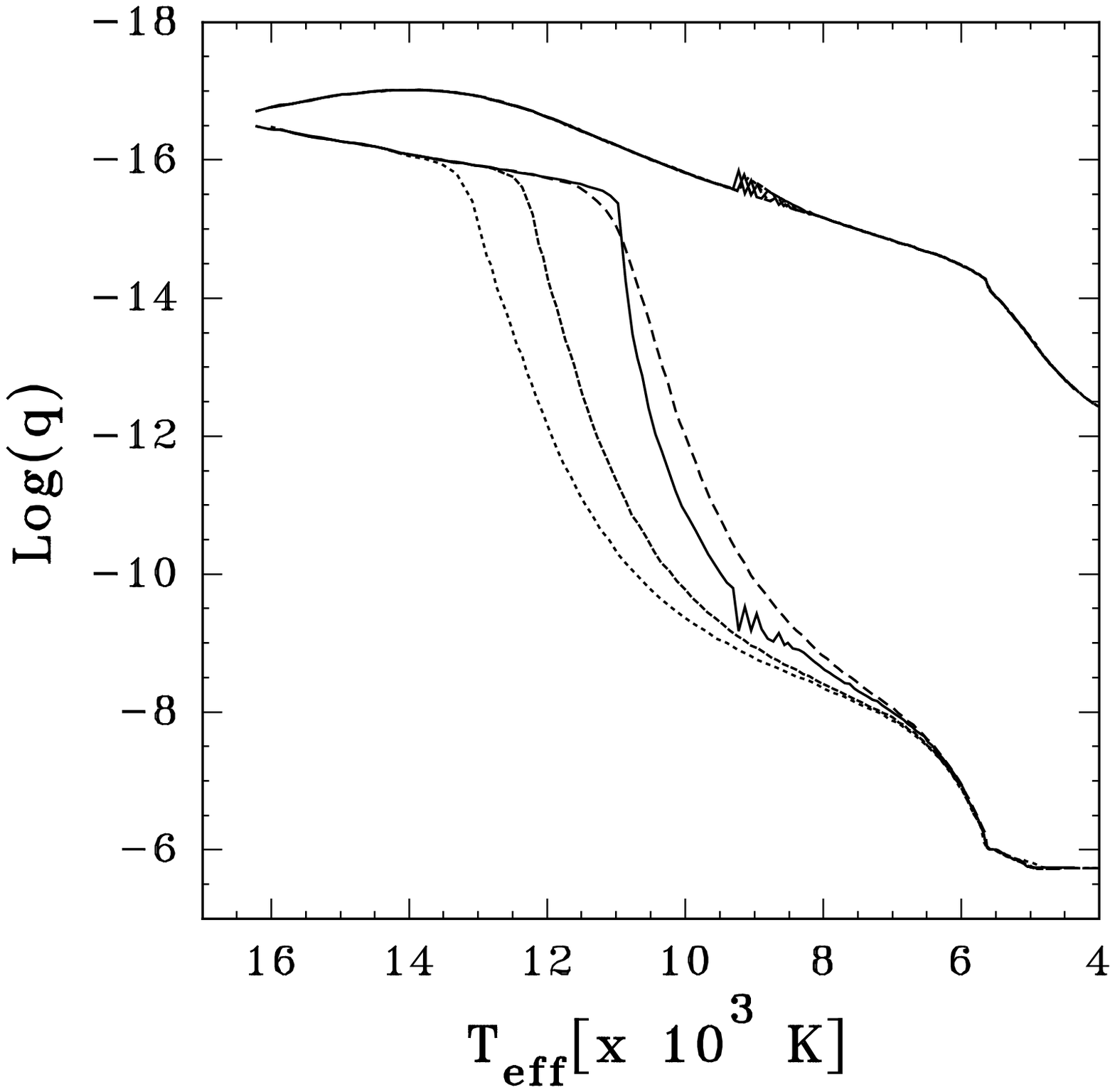}
\end{displaymath}
\caption{
Same as Fig. 4, but for a 0.5  $M_{\sun}$
DA white dwarf model.
}
\end{figure}

\begin{figure}
\epsfxsize=210pt
\begin{displaymath}
\epsfbox{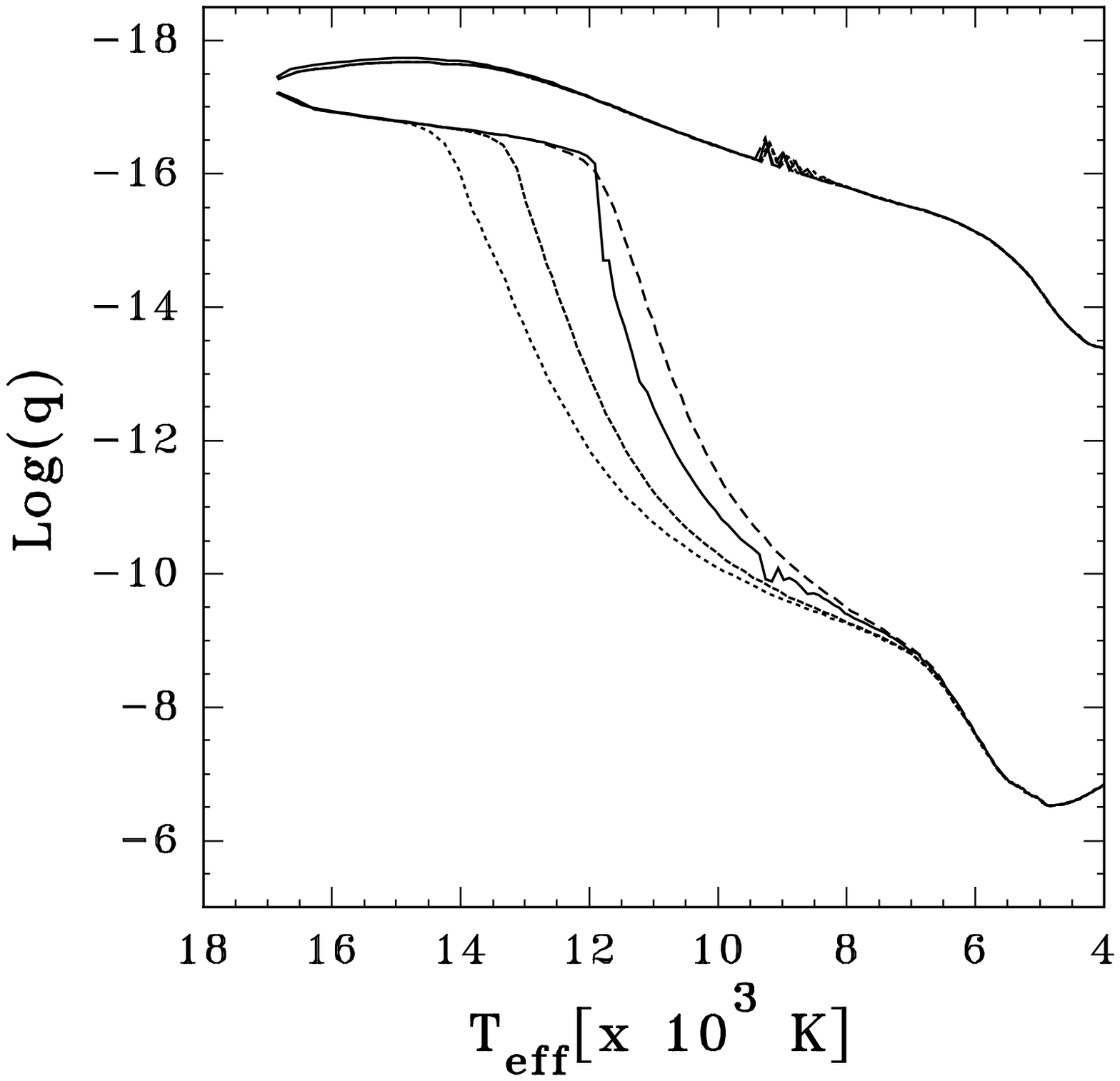}
\end{displaymath}
\caption{
Same  as  Fig. 4, but  for  a 0.8
$M_{\sun}$ DA white dwarf  model.
}
\end{figure}

\begin{figure}
\epsfxsize=210pt
\begin{displaymath}
\epsfbox{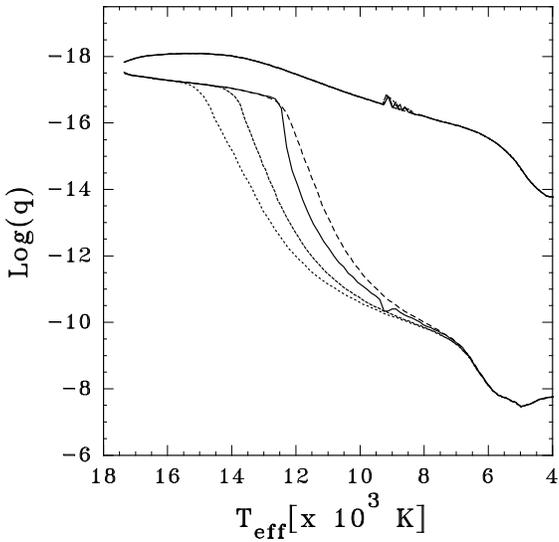}
\end{displaymath}
\caption{
Same as Fig. 4, but for a 1.0  $M_{\sun}$
DA white dwarf model.
}
\end{figure}
To clarify  better the  role played  by convective  efficiency in
determining the thermal  structure of the  outer layers, we  show
first of all  in  Fig.  8  the dimensionless
convective
temperature gradient  in terms of the  mass fraction  $q$ for the
outer layers of a 0.6  $M_{\sun}$ DA model with $T_{\rm  eff}= 10
890  K$  according  to  the  CGM  model  and  the ML1 and ML2
versions  of  MLT.  Note  the  presence  of a strong peak in
$\nabla$ at $\log q \approx  -16.1$ in the CGM case.  This arises
simply because  $z <  H_{\rm p}$  in the  outermost layers,  thus
giving rise  to   very inefficient  convection. In somewhat
deeper layers,
however, the CGM  model turns out  to be more  efficient than
ML1. This is  a result of the  fact that the  larger number of
eddies
characterizing the CGM model,  compared with  MLT, begins to
play a significant role. For $\log  q > -15.6$, both the CGM  and
ML2  models  provide  an  adiabatic  stratification and hence the
behaviour of $\nabla$  in the outermost  layers will be  critical
for determining the temperaute  profile in the envelope  and thus
the extent  of the  convection zone  of the  model. The resulting
temperature   profile   of    the   model   analysed    in   Fig.
8
is  depicted  in  Fig.  9. It is
clear that the CGM model predictions are very different from
those
of the ML1 and ML2 models. Indeed, the temperature profile cannot
be reproduced with an MLT model with a single value of $\alpha$.
\begin{figure}
\epsfxsize=210pt
\begin{displaymath}
\epsfbox{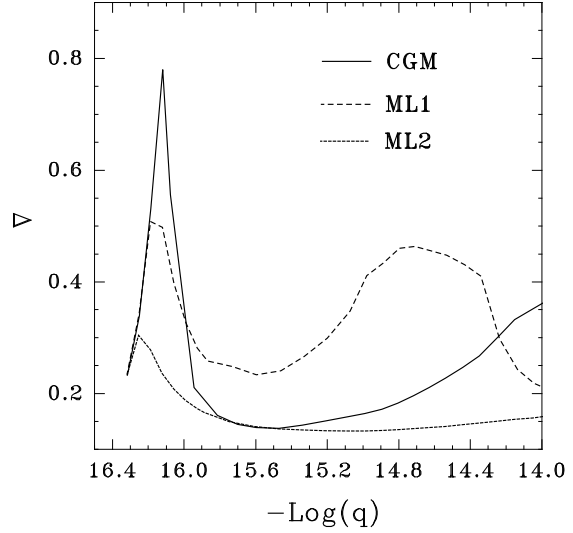}
\end{displaymath}
\caption{
The convective temperature gradient as  a
function of the mass fraction $q$ for the 0.6 $M_{\sun}$ DA white
dwarf model at $T_{\rm eff}=  10890 K$ according to the  CGM, and
to the ML1 and ML2 versions of the MLT. The fact that $l < H_{\rm
p}$  in  the  outermost  layers   leads  to  a  strong  peak   in
$\nabla$     in      the     case      of     the      CGM
model.
}
\end{figure}

\begin{figure}
\epsfxsize=210pt
\begin{displaymath}
\epsfbox{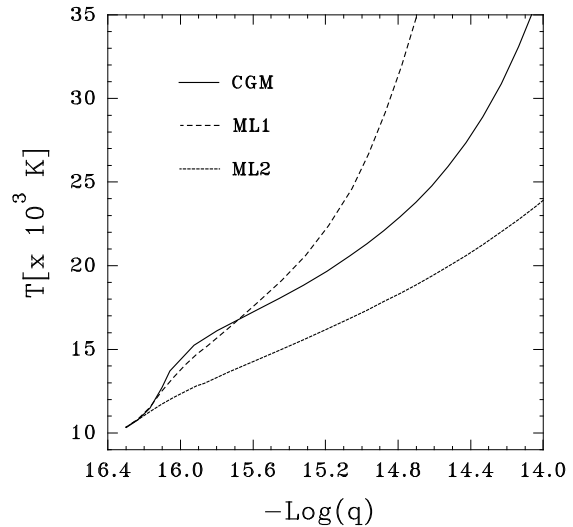}
\end{displaymath}
\caption{
The  outer layer  temperature versus  the
mass fraction $q$.  The DA model  corresponds to that  of Fig. 8.
Note that  the temperature  profile given  by the  CGM cannot  be
reproduced  with an    MLT    with    a    single    value of
$\alpha$.
}
\end{figure}

A  final  remark  about  the  convection  profiles is that at low
effective temperatures they  ultimately reach a  maximum depth,
which  is  determined  by  the  location  of  the boundary of the
degenerate core. For the 0.60 $M_{\sun}$ DA white dwarf model, in
particular,  such a depth occurs around  $\log  q  \approx -6$.
(see
Tassoul et al.  1990 for a  similar result). Accordingly,  mixing
episodes between hydrogen - rich and the underlying helium - rich
zones are possible only for values of $\log q(\rm H)$ lower  than
this  limit.  It  is   apparent  from  Figs.  4 - 7
that the mixing effective temperature for  models
with very thin hydrogen  envelopes will be strongly  dependent on
the assumed convective efficiency and not so much on the  stellar
mass.
\begin{table*}
 \centering
\begin{minipage}{140mm}
\caption{Mixing  temperatures  and  final hydrogen abundance
for different models and convection theories}
\begin{tabular}{@{}lcllc@{}}
Sequence & ML1 & ML2 & ML3 & CGM \\
0.50\ (-12) & 10,000 ($<$1e-6) & 11,240 ($<$1e-6)  & 11,870
 ($<$1e-6) & 10,400 ($<$1e-6)  \\
0.50\ (-10) & 9130 (1e-5) & 10,140 (9e-6)  & 10,700 (9e-6) &
9500 (1e-5) \\
0.50\ (-8) & 7090 (2e-3) & 7450 (1.7e-3)  & 7470 (1.7e-3) &
7220 (1.7e-3) \\
0.50\ (-6) & 5600 ($\approx$0.55) & 5605 ($\approx$0.55)  & 5610
($\approx$0.55) & 5610 ($\approx$0.55)  \\
0.60\ (-13) & 10,520 ($<$1e-6) & 11,820 ($<$1e-6) & 12,505
($<$1e-6) & 10,910 ($<$1e-6) \\
0.60\ (-12) & 10,130 ($<$1e-6) & 11,380 ($<$1e-6) & 12,000
($<$1e-6) & 10,570 ($<$1e-6) \\
0.60\ (-11) & 9680 (3e-6) & 10,840 (2e-6) & 11,450 (2e-6)
& 10,020 (3e-6) \\
0.60\ (-10) & 8950 (3e-5) & 10,090 (2e-5) & 10,650 (2e-5) &
9450 (2e-5) \\
0.60\ (-8) & 6710 (7.3e-3) & 6900 (6.4e-3) & 6850  (6.8e-3) &
6780 (6.6e-3) \\
0.60\ (-6) & 5330 ($\approx$0.8) & 5360 ($\approx$0.8) & 5380
($\approx$0.8) & 5360 ($\approx$0.8) \\
0.70\ (-12) & 10,160 (1e-6) & 11,390 (1e-6) & 12,100  (1e-6) &
10,640 (1e-6) \\
0.70\ (-10) & 8980 (8e-5) & 9890 (6e-5) & 10,330  (6e-5) &
9320 (7e-5) \\
0.70\ (-8) & 6410 (0.027) & 6530 (0.025) & 6530 (0.026)  &
6500 (0.025) \\
0.80\ (-12) & 10,250 (2e-6) & 11,500 (1e-6) & 12,020 (1e-6)  &
10,760 (1e-6) \\
0.80\ (-10) & 8740 (2.2e-4) & 9740 (1.7e-4) & 10,010 (1.6e-4)  &
9180 (1.8e-4) \\
0.80\ (-8) & 6320 (0.1) & 6330 (0.1) & 6350 (0.1)  &
6350 (0.1) \\
0.90\ (-12) & 10,350 (4e-6) & 11,470 (3e-6) & 12,050 (3e-6)  &
10,830 (3e-6) \\
0.90\ (-10) & 8560 (6.5e-4) & 9320 (5e-4) & 9490 (4.6e-4)  &
9120 (5e-4) \\
0.90\ (-8) & 6170 ($\approx$0.4) & 6200 ($\approx$0.4) &  6220
($\approx$0.4)  & 6230 ($\approx$0.4) \\
1.0\ (-12) & 10,290 (1e-5) & 11,560 (7e-6) & 11,980 (7e-6)  &
10,800 (8e-6) \\
1.0\ (-10) & 8240 (2e-3) & 8810 (1.6e-3) & 8950 (1.6e-3)  &
8610 (1.6e-3) \\
1.0\ (-8) & 5950 ($\approx$0.8) & 5970 ($\approx$0.8) &  5990
($\approx$0.8)  & 6000 ($\approx$0.8)
\end{tabular}
\medskip

Each  sequence  is  denoted  by  the  stellar mass (in solar mass
units) and the value of $\log\ q_ H$. The numbers in  parentheses
next to each temperature give the hydrogen surface content  after
mixing.
\end{minipage}
\end{table*}

\begin{figure}
\epsfxsize=210pt
\begin{displaymath}
\epsfbox{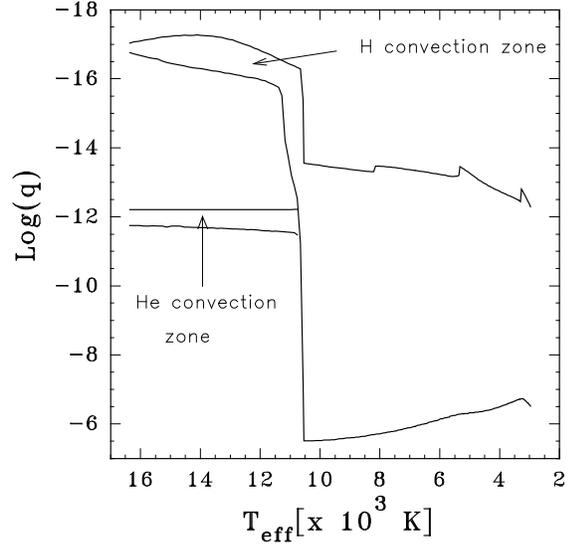}
\end{displaymath}
\caption{
The location of  the top and the  base of
the  outer  convection  zones  expressed  in  terms  of  the mass
fraction  $q$  versus  the  effective  temperature.  The  results
correspond to a  0.6 $M_{\sun}$ DA  white dwarf model  with $\log
q(\rm H)= -12$ according to  the CGM model. Note the  presence of
two convection zones at high $T_{\rm eff}$ which ultimately merge
at  $T_{\rm}  \approx  10500  K$,  thus  giving rise to a drastic
change  of  the  surface  chemical  composition from a hydrogen -
dominated  to  a  helium  -  dominated  one.
}
\end{figure}

To put this assertion on a more quantitative basis,
we  list  in  Table  1  the  effective  temperatures at which the
hydrogen surface  abundance begins to decrease  appreciably as  a
result  of  convective  mixing, together  with the final hydrogen
abundances (for models with thick hydrogen layers, mixing process
takes place gradually in a finite range of effective temperature;
in that  case we  only give  approximate values  of the  hydrogen
abundance).  Clearly,   for  models   with  very   thin  hydrogen
envelopes,  convective  mixing  drastically  modifies  the
surface
composition from a hydrogen -  dominated to a helium -  dominated
one. This behaviour can be understood directly  by  examinig
Fig.
10, which shows the evolving convection zones of the
$0.6M_{\sun}$
model with $\log q(\rm H) = -12$ according to the CGM convection.
At $T_{\rm  eff} \approx  10600 K$,  the merging  of the hydrogen
convection zone  with the  helium convection  zone (located  just
below the hydrogen/helium transition  region) causes the base  of
the convection  zone to  reach very  deep helium  layers quickly,
giving  rise  to  an  almost  complete  dilution  of the hydrogen
content. From  then on,  the subsequent  evolution corresponds to
that of a DB model (see Fig. 11 of Benvenuto \& Althaus 1997).

In order to estimate  approximate effective temperatures for  the
theoretical blue edge of  DA instability strip we  employ thermal
time - scale  arguments  $\tau_{\rm  th}$  for  our evolving
models.
Several past studies  (Cox 1980, Winget  et al. 1982,  Tassoul et
al.  1990  and  references  cited  therein)  have  shown that the
behaviour of pulsational instabilities  in white dwarf stars  can
be understood in terms of $\tau_{\rm th}$ of the driving  regions
defined by

\begin{equation}
\tau_{\rm th}= \int^{q_{\rm bc}}_{0} {{C_{\rm V}\ T}\over{L}}
M_{\rm *}\ dq
\end{equation}

In  Eq. 6,  $C_{\rm  V}$  is  the  specific heat at
constant volume,  $L$ and  $M$ are,  respectively, the luminosity
and the mass of the  model, and the subscript ``bc''  corresponds
\begin{table}
\caption{Theoretical  blue-edge  effective temperatures
versus stellar mass  and convection theory for our DA white
dwarf}
\begin{tabular}{@{}lcc@{}}
Theory of convection & Mass ($M/M_{\odot}$) & $T_{eff}\ (K)$
\\
CGM & 0.50 &  10,710   \\
ML1 &   '' &  10,400   \\
ML2 &   '' &  11,630   \\
ML3 &   '' &  12,310   \\
CGM & 0.60 &  10,970   \\
ML1 &   '' &  10,600   \\
ML2 &   '' &  11,850   \\
ML3 &   '' &  12,530   \\
CGM & 0.70 &  11,210   \\
ML1 &   '' &  10,800   \\
ML2 &   '' &  12,050   \\
ML3 &   '' &  12,710   \\
CGM & 0.80 &  11,430   \\
ML1 &   '' &  10,980   \\
ML2 &   '' &  12,240   \\
ML3 &   '' &  12,890   \\
CGM & 0.90 &  11,650   \\
ML1 &   '' &  11,160   \\
ML2 &   '' &  12,430   \\
ML3 &   '' &  13,070   \\
CGM &  1.0 &  11,870   \\
ML1 &   '' &  11,350   \\
ML2 &   '' &  12,620   \\
ML3 &   '' &  13,250
\end{tabular}
\medskip
\end{table}

to the location of the bottom of the superficial convection zone.
In particular, the blue edge of the instability strip (the  onset
of  pulsations)  corresponds   approximately  to  the   effective
temperature at which $\tau_{\rm th}$ becomes comparable to 100 s,
which  for  a  white  dwarf  is  of  the  order  of  the shortest
observable g - mode periods. In this connection, we list in Table
2 (see also Fig. 11) the effective temperature of
our  theoretical  blue  edges  of  the  ZZ Ceti instability strip
according  to  the   different  stellar  masses   and  convection
treatments  we   have  considered.   Note,  as   found  by  other
investigators (e.g.  Tassoul et  al. 1990  and Bradley  \& Winget
1994) the dependence of the blue edge temperature on the  stellar
mass and, more importantly,  on the assumed convective
efficiency.
In particular, our  MLT blue edges  are consistent with  those of
Bradley  \&  Winget  (1994)  obtained  on  the  basis of detailed
pulsation calculations.  Clearly the  CGM theory  predictions are
intermediate  to  the   ML1  and  ML2   results.  In  Fig.  11,
we  have  also  included  the  theoretical result
obtained by Gautschy et al.  (1996) for a $0.6 M_{\sun}$  ZZ Ceti
model and  the observational  data of  Bergeron et  al. (1995) as
well (we picked out those  ZZ Ceti stars analysed by  Bergeron et
al. which,  for a  given stellar  mass, are  characterized by the
highest temperature).

With regard to  the inclusion of  the observational data  in this
figure,  some  comments  seem  to  be  appropriate at this point.
As
mentioned earlier,  the effective  temperature of  the stars that
define the {\it observational} blue edge is computed, as  usual,
by employing the emergent  spectrum of a model  atmosphere, which
in the  case of  DA white  dwarfs is
\begin{figure}
\epsfxsize=210pt
\begin{displaymath}
\epsfbox{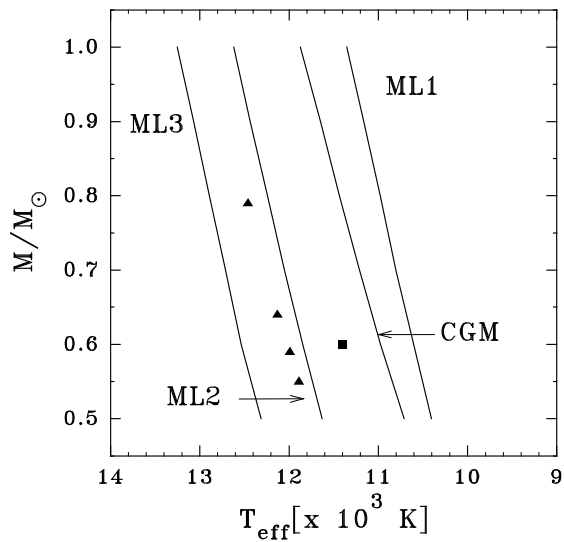}
\end{displaymath}
\caption{
The dependence  of theoretical blue  edge
temperature on the stellar mass for CGM, and for the ML1, ML2 and
ML3  versions  of  the  MLT.  The  filled triangles represent the
observations of Bergeron et al.  (1995) for (from top to  bottom)
G226-29, G185-32, R548 and G238-53. The filled square corresponds
to the  blue edge  temperature obtained  by Gautschy  et al.
(1996)
from hydrodynamical simulations  of convection. Note
that both the CGM model and the hydrodynamical simulation tend to
point   towards   a   cooler   blue   edge   than   the  observed
one.
}
\end{figure}
strongly dependent  on the
prescription adopted  for the  treatment of  convection (see e.g.
Bergeron  et  al.  1995).  Thus,  in  order  to  perform a self -
consistent comparison with observations, we should employ a model
atmosphere computed considering  the CGM theory.  In view of  the
fact that such models are still not available, any comparison  of
our theoretical  results with  observation should  be taken  with
caution.

In spite of  the above warnings,  it is remarkable  that both the
CGM  theory  and  Gautschy  et  al.'s (1996)  predictions  tend
to point
towards a cooler blue  edge than observations. In  particular, on
the  basis  of   very  detailed  hydrodynamical   simulations  of
convection in  a $0.6  M_{\sun}$ ZZ  Ceti model,  Gautschy et al.
derived an effective temperature between 11400 K and 11800 K  for
the blue edge. As far as the blue edge dependence on the  stellar
mass is concerned, the MLT and CGM theory predictions are  rather
similar,  spanning  $\approx  1000  K$  over the range of stellar
masses we considered. Note that dependence of the blue edge on
the  stellar  mass  predicted  by  both theories of convection is
rather similar to that shown by observations.

\section{Summary}

In this paper we compute the structure and evolution of carbon  -
oxygen white dwarf models  with hydrogen envelopes (DA  type). We
consider stellar masses ranging from M=0.5  to  $M=1.0\
M_{\sun}$ at intervals of $M=0.1 M_{\sun}$, and we treat the
mass
of the hydrogen  and helium envelopes  as free parameters  within
the range $10^{-13} \leq  M_{\rm H}/M \leq 10^{-4}$  and $10^{-6}
\leq M_{\rm  He}/M \leq  10^{-2}$, respectively.  The models were
evolved from the intermediate effective temperature stage down to
$\log{(L/L_{\sun})} = -5$.

The calculations were made  with a white dwarf  evolutionary code
including updated radiative and conductive opacities and neutrino
emission rates, and very detailed equations of state for hydrogen
and   helium   plasmas.   We   also   include   the   effects  of
crystallization, convective mixing and hydrogen burning (pp chain
and CNO bi - cycle). The most important feature of our  study,
however,
is that we  treat the energy  transport by convection  within the
formalism of  the so  - called  full - spectrum  turbulence
theory,
which constitutes an improvement over most previous studies of
DA white dwarf  evolution. In particular,  we employ a  new model
based on such theory  developed by Canuto, Goldman  \& Mazzitelli
(1996)  (CGM).  This  new  model  includes  the  full spectrum of
eddies, has  no free  parameter and  computes the  rate of energy
input self  - consistently.  In the  white dwarf  domain, the CGM
models has been  recently shown (Althaus  \& Benvenuto 1997b)  to
provide a  good fit  to new  observational data  of pulsating DB
objects. Finally, for  the sake of  comparison, we also  consider
the  most  common  parametrizations  of  the mixing length theory
(MLT) usually employed in this kind of studies.

In  agreement  with  previous  studies,  we  find that very thick
hydrogen layers substantially modify  the surface gravity of  the
models and  that the  importance of  nuclear burning  is strongly
sensitive  to  the  excat  value  of  the hydrogen layer mass. In
particular,  we  find  that  for  the  0.6  $M_{\sun}$ model with
$M_{\rm H}= 9 \times 10^{-5} M_{\sun}$ the relative  contribution
of  hydrogen  burning  at  low  luminosities remains always below
9 per cent, while for the same  model but with $M_{\rm H}=  1.2
\times
10^{-4}  M_{\sun}$, the  hydrogen  burning  contribution  rises
up to
$\approx 18$ per cent.

One of  our main  interest in  this work  has been  to study  the
evolution of ZZ Ceti models with the aim of comparing the CGM and
MLT predictions. In this connection, we find that the temperature
profile given by  the CGM model  is markedly different  from
that of the ML1 and ML2 models, and it cannot be reproduced by
any
MLT model  with a  single value  of $\alpha$.  The evolving outer
convection zone also behaves  differently in  both theories.
In
particular, the thickness of the convection zone in the CGM model
begins to  increase at  a given  effective temperature  much more
steeply than in any of the MLT versions and remains  intermediate
to those with ML1 and ML2 convection.

We have also computed approximate effective temperatures for  the
theoretical blue edge  of DA instability  strip by using  thermal
time - scale arguments for our evolving DA models. In this
context,
we find that  the effective temperature  of the theoretical  blue
edges of the  ZZ Ceti instability  strip depends strongly  on the
stellar  mass  and  more  importantly  on  the assumed convective
efficiency, in  agreement with  previuos studies.  In particular,
our MLT blue edges are consistent with those of Bradley \& Winget
(1994) obtained on the basis of detailed pulsation  calculations.
We find  that the  CGM theory  predicts blue  edges
cooler   (by   $\approx   1000   K$)   than  the  observed  ones.
Specifically, we find an  effective temperature of $11000  K$ for
the blue edge of a 0.6 $M_{\sun}$ DA model. It is remarkable that
recent  non - adiabatic  pulsation  calculations  based on
numerical
simulations of convection tend also to indicate a somewhat cooler
blue  edge.  However,  we  remaind  the  reader that the physical
ingredients  we  employed  in  the  present  paper  are {\it not}
consistent  with  those  of  the  stellar atmosphere calculations
employed by Bergeron et al. (1995), particularly in the treatment
of  convection.  Although  a  difference  of  1000  K between the
effective temperature of the observed and theoretical blue  edges
for the ZZ  Ceti instability strip  seems to be  large, we should
wait for model atmospheres also computed in the frame of the full
spectrum  turbulence  theory.  Only  after  such  models   become
available we  shall be  able to  gauge the  actual importance of
such a discrepancy.

Detailed tabulations of the evolution of our DA models, which are
not       reproduced       here,       are      available      at
http://www.fcaglp.unlp.edu.ar/ $\sim$althaus/

\section*{acknowledgments}

We thank  Professor R.  Stothers for  providing us  with material
before   its   publication.   We   also   appreciate   e  -  mail
communications with  H.-G. Ludwig.  This work  has been partially
supported by the  Comisi\'on de Investigaciones  Cient\'{\i}ficas
de  la  Provincia  de  Buenos  Aires,  the  Consejo  Nacional  de
Investigaciones Cient\'{\i}ficas y T\'ecnicas (Argentina) through
the  Programa   de  Fotometr\'{\i}a   y  Estructura   Gal\'actica
(PROFOEG) and the University of La Plata.

{}


\begin{thebibliography}{99}


\bibitem{} Alexander  D. R.,   Ferguson  J. W., 1994, ApJ,
437, 879

\bibitem{} Althaus L.  G.,  Benvenuto  O. G., 1996,  MNRAS,
278, 981

\bibitem{} Althaus  L. G.,   Benvenuto  O. G., 1997a, ApJ,
477, 313

\bibitem{} Althaus L. G.,   Benvenuto O. G., 1997b, MNRAS,
288, L35

\bibitem{} Benvenuto  O. G.,   Althaus  L. G.,  1995, Ap\&SS,
234, 11

\bibitem{} Benvenuto  O. G.,   Althaus  L. G.,  1996, ApJ,
462, 364

\bibitem{} Benvenuto O. G.,  Althaus L. G., 1997, MNRAS, 288,
1004

\bibitem{} Benvenuto O. G.,  Althaus L. G., 1998, MNRAS, 293, 177

\bibitem{} Bergeron P., Wesemael F., Fontaine G., 1992a, ApJ,
387, 288

\bibitem{} Bergeron P., Saffer R. A., Liebert J., 1992b, ApJ,
394, 228

\bibitem{} Bergeron P., Wesemael F., Lamontagne R., Fontaine
G., Saffer R. A., Allard N. F., 1995, ApJ, 449, 258

\bibitem{} B\"{o}hm - Vitense E., 1958, Z. Astrophys., 46, 108

\bibitem{} Bradley P. A., 1996, ApJ, 468, 350

\bibitem{} Bradley P.  A.,  Winget  D. E., 1994,  ApJ,
421, 236

\bibitem{} Canuto V. M., 1996, ApJ 467, 385

\bibitem{} Canuto V. M., Christensen  - Dalsgaard J.,
1997 Annu. Rev. Fluid Mech., in press

\bibitem{} Canuto V.  M.,  Mazzitelli  I., 1991, ApJ,  370,
295 (CM)

\bibitem{} Canuto V.  M.,  Mazzitelli  I., 1992, ApJ,  389,
724 (CM)

\bibitem{} Canuto V. M., Goldman I.,  Mazzitelli I., 1996,
ApJ, 473, 550 (CGM)

\bibitem{} Caughlan G. R.,  Fowler W. A., 1988, Atomic  Data
and Nuclear Data Tables, 40, 290

\bibitem{} Cox J. P., 1980, Theory of Stellar Pulsations.
Princeton University Press, New Jersey

\bibitem{} Cox A. N.,  Stewart J., 1970, ApJs, 19, 261

\bibitem{} D'Antona F.,  Mazzitelli I., 1979, A\&A, 74, 161

\bibitem{} D'Antona  F., Mazzitelli  I., 1991,  in Michaud  G.,
Tutukov  A.,  eds,  IAU  Symp.  145,  Evolution  of  Stars:   the
Photospheric Abundance Connection. Kluwer, Dordrecht, p. 339

\bibitem{} D'Antona F.,  Mazzitelli I., 1994, ApJs, 90, 467

\bibitem{} D'Antona F.,  Mazzitelli I., Gratton R. G., 1992,
A\&A, 257, 539

\bibitem{} D'Antona F., Caloi V., Mazzitelli I., 1997, ApJ,
477, 519

\bibitem{} Dolez N.,  Vauclair G., 1981, A\&A, 102, 375

\bibitem{} Fontaine G., Van Horn H. M., 1976, ApJs, 31, 467

\bibitem{} Fontaine G., Brassard P., Bergeron P., Wesemael  F.,
1996, ApJ, 469, 320

\bibitem{} Fontaine G., Brassard P., Wesemael F., Tassoul M.,
1994, ApJ, 428, L61

\bibitem{} Gautschy A., Ludwig H.-G., Freytag B., 1996, A\&A,
311, 493

\bibitem{} Hubbard W. B., Lampe M., 1969, ApJs, 18, 297

\bibitem{} Iben I. Jr.,  Tutukov A. V., 1984, ApJ, 282, 615

\bibitem{} Iglesias C. A.,   Rogers F. J., 1993, ApJ,  412,
752

\bibitem{} Koester D.,  Sch\"{o}nberner D., 1986, A\&A, 154,
125

\bibitem{} Koester D., Allard N. F., Vauclair G., 1994, A\&A,
291, L9,

\bibitem{} Kupka F. 1996, in Adelman S. J., Kupka F., Weiss W. W.
eds, Model Atmospheres and Spectrum Synthesis, ASP, 108, 73

\bibitem{} Ludwig H.-G., Jordan S., Steffen M., 1994, A\&A,
284, 105

\bibitem{} McGraw J. T., 1979, ApJ, 229, 203

\bibitem{} Marsh M. C. et al. , 1997, MNRAS, 286, 369

\bibitem{} Mazzitelli I., 1995, in Werner K., Koester D., eds,
White Dwarfs, Springer, Berlin, 58

\bibitem{} Mazzitelli I.,  D'Antona F., 1986, ApJ, 308, 706

\bibitem{} Mazzitelli I.,  D'Antona F., 1991, in  Vauclair G.,
Sion E. M., eds, Seventh  European  Workshop  on  White Dwarfs
(NATO ASI Series). Kluwer, Dordrecht, p. 305

\bibitem{} Monteiro M. J. P. F. G., Christensen-Dalsgaard
J., \& Thompson M. J., 1996, A\&A, 307, 624

\bibitem{} Patern\`o  L.,  Ventura R., Canuto V. M, Mazzitelli
I., 1993, ApJ, 402, 733

\bibitem{} Saumon D., Chabrier G.,   Van Horn H. M., 1995,
ApJs, 99, 713

\bibitem{} Stothers R. B., Chin C.-W., 1995, ApJ, 440, 297

\bibitem{} Stothers R. B., Chin C.-W., 1997, ApJ, 478, L103

\bibitem{} Tassoul M.,  Fontaine G.,   Winget D. E., 1990,
ApJs, 72, 335

\bibitem{} Thejll P.,  Vennes S., Shipman H. L.,  1991,
ApJ, 370, 355

\bibitem{} Wagoner R. V., 1969, ApJs, 18, 247

\bibitem{} Wallace  R. K.,  Woosley S.  E.,   Weaver T. A.,
1982, ApJ, 258, 696

\bibitem{} Weidemann  V.,  Koester D., 1984, A\&A, 132, 195

\bibitem{} Wesemael  F., Bergeron  P., Fontaine  G., Lamontagne
R.,  1991,  in  Vauclair  G.,  Sion  E. M., eds, Seventh European
Workshop on White Dwarfs (NATO ASI Series). Kluwer, Dordrecht, p.
159

\bibitem{} Winget D. E., Van Horn H. M., Tassoul M., Hansen
C. J., Fontaine G., Carroll B. W., 1982, ApJ, 252, L65

\bibitem{} Winget D. E., et al.,  1994, ApJ, 430, 839


\end{thebibliography}
\end{document}